\documentclass[pre,twocolumn,showpacs,superscriptaddress]{revtex4}

\usepackage{amsmath,amssymb,graphicx}

\begin{document}

\title{Single file diffusion in macroscopic Wigner rings}
\date{\today}
\author{Gwennou Coupier}\author{Michel Saint Jean}\email{michel.saintjean@paris7.jussieu.fr}\author{Claudine Guthmann}
 \affiliation{Laboratoire
Mati\`ere et Syst\`emes Complexes, UMR 7057 CNRS \& Universit\'e
Paris~7 - 140 rue de Lourmel, F-75015 Paris, France}

\begin{abstract}
The single file diffusion in a circular channel of millimetric
charged balls is studied. The evolution in time of the mean square
displacement is shown to be subdiffusive, but slower than the
power-like $t^{1/2}$ behavior observed in circular colloidal systems
or predicted in one-dimensional infinite systems.
\end{abstract}
 \maketitle

\section{Introduction}
\label{intro}

Diffusion of particles in a single channel where the mutual
exchanges are forbidden is known as single file diffusion (SFD).
This physical situation is encountered in various fields, for
instance 1D hopping conductivity, ion transport in biological
membranes or molecules channeling in zeolithes. In these geometries
with low dimensionality, the particles in repulsive interaction are
not able to cross each other, and the particles remain correlated
even at long times.

Such systems were generally modeled by an infinite set of particles
with hard interaction and diffusing on a
line~\cite{harris65,vanbeijeren83}: the mean square displacement
$\Delta x^2$ has been proved to grow at long times as $t^{1/2}$.
More recently, it was shown that this behavior is also encountered
in overdamped systems with arbitrary repulsive interactions,
providing the correlation length between the particles is of finite
range~\cite{kollmann03}. On the other hand, although the main
theoretical results for single file diffusion were obtained many
years ago, experiments displaying such subdiffusive behavior are
lacking and the obtained results are often conflicting. For
instance, NMR studies~\cite{nivarthi94,hahn98} and quasi elastic
neutron scattering experiments~\cite{nivarthi94,jobic97} on organic
molecules in porous materials either conclude to subdiffusive
transport~\cite{nivarthi94,hahn98} or to classical
diffusion~\cite{nivarthi94,jobic97}, with apparently the same
experimental conditions.
\begin{figure}
\resizebox{0.8\columnwidth}{!}{\includegraphics{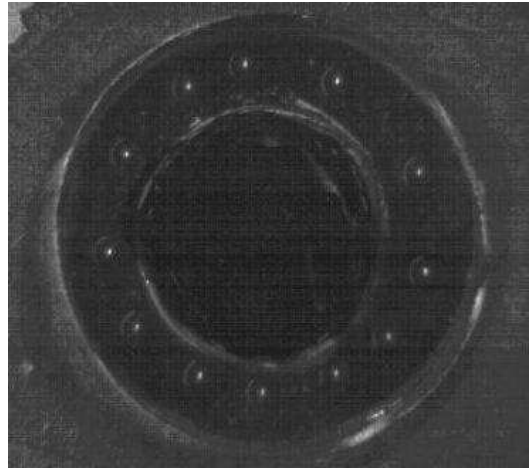}}
\caption{Photography of the experimental circular channel with
$N=12$ balls inside.} \label{fig:fig0}
\end{figure}
More recently, the movement of colloidal particles in a circular
channel has been studied. This channel is obtained by
photolithography~\cite{wei00,lin02} or optical tweezer circular
trapping~\cite{lutz04_2}. While classical diffusion is always
observed for short times, the long time behavior is not as well
clearly analyzed. Some authors suggest it grows as $t^{1/2}$ like
theoretical SFD processes for infinite systems~\cite{wei00,lutz04_2}
whereas others
 explain this subdiffusion by hydrodynamic coupling
effects~\cite{lin02}. On the other hand, the theoretical study made
in Ref.~\cite{vanbeijeren83} for a 1D line of hard-core interacting
particles with periodic boundary conditions shows that the mean
square displacement grows linearly, as for a free diffusion.

In order to rule out these ambiguities, we present here the
diffusion of macroscopic charged metallic balls electrostatically
interacting and moving in a circular channel whose width forbids any
crossing, while a mechanical shaking induces an effective
temperature. This system presents many advantages : it gives the
opportunity to study the diffusion at very long times, it allows to
suppress any hydrodynamic effect and, if necessary, to tune the
interacting forces. In this experiment, the inter-particle
interaction is similar to $K_0$ inter-vortices interaction in
superconductors~\cite{galatola05}. This interaction differs from the
dipolar interaction of colloids or from the hard interaction
introduced in many theories, thus we can evaluate the influence of
the interaction characteristics on the SFD behaviors. Note that no
hard-core collisions are observed in our system.

Our main result is to exhibit subdiffusive behavior slower than the
$t^{1/2}$ behavior predicted by theory and mentioned in colloidal
systems.

In Sec.~\ref{sec:validation} we present the experimental set-up and
prove its ability to describe diffusion processes by studying two
basic cases. To identify the effects of the circular confinement on
the diffusion in order to distinguish them from the inter-particle
interaction contributions and to determine the best conditions to
obtain the required 1D movement, we have studied the diffusion of a
single ball in the channel. The obtained results are reported in
Sec.~\ref{sec:oneball}. Sec.~\ref{sec:nballs} is devoted to the
diffusion of $N=$ 12 and 16 interacting particles and to the
comparison with the behaviors previously presented in literature, in
particular those obtained with colloids.

\section{Experiment validation}
\label{sec:validation}

In this experiment, millimetric stainless steel balls (of radius $R
= 0.4$ mm and weight $m=2.15$ mg) are located on the doped-silicon
 bottom electrode of an horizontal plane condenser (of height $h=1.5$
mm). A metallic frame intercalated between the two electrodes and in
contact with the bottom one confines the balls in a circular
channel. Its external diameter and its width are respectively 10 mm
and 2 mm (Fig.\ref{fig:fig0}). This forbids any crossing between the
balls. In order to charge the balls, a tunable voltage $V_c$ of
about 1kV is applied to the top electrode. We then get a system
where the interaction between the balls as well as the confining
potential are easily adjusted. The interaction potential $V(r)$
between two balls has been shown to be well described through a
modified Bessel function of the second kind : $V(r)=A
K_0(r/\lambda)$ where the screening length $\lambda$ is about
$0.3h=0.5$ mm and independent from $V_c$, as discussed in details in
Ref.~\cite{galatola05}. Direct capture of the position of the balls
is made by a camera placed above the experimental device. The
typical time between two snapshots is selected between 15 ms and 150
ms and series of 10,000 images were recorded. The individual
trajectories of these particles are thus directly determined over
very long times (comparing with the typical relaxation time that
will be determined in the following).

To introduce the thermal noise, the whole cell is fixed on
loudspeakers supplied by a white noise voltage. Thanks to the
friction between the bottom electrode and the balls, a erratic and
non spatially correlated movement is conveyed to the latter. It has
been previously shown that the ensemble of balls obey Boltzmann
statistics, where the shaking amplitude stands for the
temperature~\cite{coupier05}. The effective temperature is thus
determined in situ by measuring the mean square position of a single
ball confined by a circular frame.

The diffusion of the particles is characterized by the evolution
with time of their mean square displacements (m.s.d.) given by, if
the considered coordinate is $x$ :
\begin{equation}
\Delta x^2(t)=\langle (x(t)-\langle x(t)\rangle)^2\rangle,
\end{equation}

where the brackets $\langle\,\rangle$ denote ensemble averaging. In
practice, we get from the trajectory of one ball a set of
statistically independent trajectories by shifting the time origin.
We emphasize that such a method implies unavoidable averaging over
the distribution of initial conditions, which can lead to mean
square displacements slightly different than the usual ones (see the
appendix). Furthermore, when considering the orthoradial coordinate
$\theta$ (in radians) we will consider the cumulated angle and not
the modulo $2\pi$ angle, since we want to consider the ring as an
easy realization of a 1D line. As in
Ref.~\cite{wei00,lin02,lutz04_2}, the total orthoradial displacement
of a ball is taken into account, without any substraction of a mean
displacement of the system.

In order to evaluate the ability of our experimental set-up to
explore diffusion processes, two simple cases for which the
diffusion behaviors are well established have been studied.

\begin{figure}
\resizebox{\columnwidth}{!}{\includegraphics{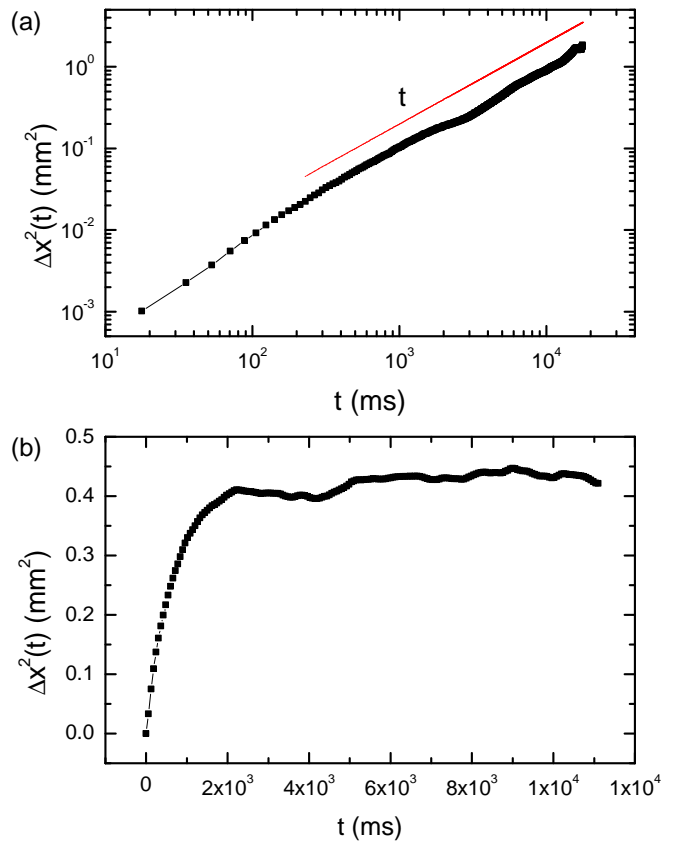}}
\caption{\label{fig:fig1} (a) Typical mean square displacement of a
free ball (log scale) ; (b) typical mean square displacement of a
ball trapped in a well constituted by a circular frame of radius 5
mm. The applied voltage $V_c$ is 1000 V.}
\end{figure}

The free diffusion of a ball moving on the bottom electrode without
any confinement and any applied electrostatic potential was first
studied. Whatever the effective temperature, the m.s.d. $\Delta x^2
(t)$, $x$ being the ball abscissa, presents the same typical time
variation, presented on Fig.~\ref{fig:fig1}(a). This m.s.d. varies
linearly with time at long times, whereas the short time behavior
exhibits a $t^\beta$ power increase with $1<\beta<2$. These
behaviors are in agreement with a free diffusion described by the
classical Langevin equation which predicts $\Delta x^2 (t) \simeq
2Dt$ at long times where $D$ is the diffusion constant. Since the
temperature is independently measured, the effective damping
coefficient $\gamma$ can be determined from the
fluctuation-dissipation theorem  $m\gamma D= k_B T$. The related
relaxation time $\tau_R= 1/\gamma$  which marks the crossover
between the short and long time regimes is found to be equal to
about 100 ms, which is coherent with a rough estimation on the curve
of Fig.~\ref{fig:fig1}(a).

This value which is quite close to the smallest snapshot time
explains why the theoretical behavior in $t^2$ predicted for very
short times ($t\ll\tau_R = 1/\gamma$) may not be observed, the
behavior observed at short times corresponding actually to the
progressive transition between a $t^2$ regime to a linear one (this
analyze is confirmed by the measured mean square velocity which is
always constant in time as expected in the Langevin theory for  $t >
\tau_R$).

The second test was to follow the diffusion of a particle confined
in a well created by a circular frame (of radius 5 mm). In such a
case, the ball moves in a parabolic confining potential whose
stiffness $K$ can be determined from the radial distribution of the
ball~\cite{coupier05}. Fig.~\ref{fig:fig1}(b) presents a typical
radial m.s.d. of a ball in such potential. After a rapid increase,
it reaches a constant value which varies linearly with the effective
temperature as expected. These behaviors are those predicted in a
Langevin formalism in which a parabolic potential models the
confinement. At short times ($t < (K/m)^{1/2}$), the ball begins to
follow a free diffusion since the particle has no time to explore
the whole confining potential whereas at long times ($t >
(K/m)^{1/2}$) the experimental m.s.d. is dependent on the
confinement and reaches $2k_BT/K$ (see the appendix). In the same
time, the angular m.s.d grows linearly with time, which
characterizes a free diffusion along the orthoradial direction.

The accordance between the theoretical behavior and the observed one
confirms the ability of our experimental set-up to give relevant and
self coherent parameters about particle diffusion and that the
Langevin formalism describing Brownian motion can be used.

\section{Diffusion of a single ball in a circular channel.}
\label{sec:oneball}

In our experiments, the interacting balls move in a electrostatic
confining potential which looks like a  circular gutter. In order to
distinguish possible effects due to this channel geometry from those
resulting from the inter-particle interaction, we have preliminary
studied the diffusion of a single ball in such a circular channel.
In particular, the radial confining potential has been studied for
different applied electrostatic potentials in order to get the best
opportunities for a quasi 1D movement. In the same time, the
coupling of the small radial displacement with the orthoradial
movement inherent to the circular movement has also been evaluated.

\begin{figure}
\resizebox{\columnwidth}{!}{\includegraphics{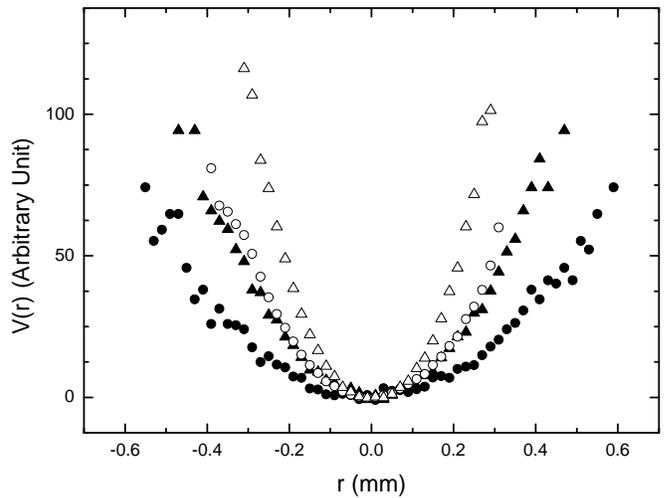}}
\caption{Confining radial potential in the circular channel obtained
from the radial distribution of the balls, for two different numbers
of balls and two different applied voltage $V_c$ : 1 ball, $V_c=800$
V ($\bullet$); 1 ball, $V_c=1000$ V ($\blacktriangle$); 12 balls,
$V_c=800$ V ($\circ$); 12 balls, $V_c=1000$ V ($\vartriangle$). $r$
is measured relatively to the mean radial position.}
\label{fig:fig2}
\end{figure}

\begin{figure}
\resizebox{\columnwidth}{!}{\includegraphics{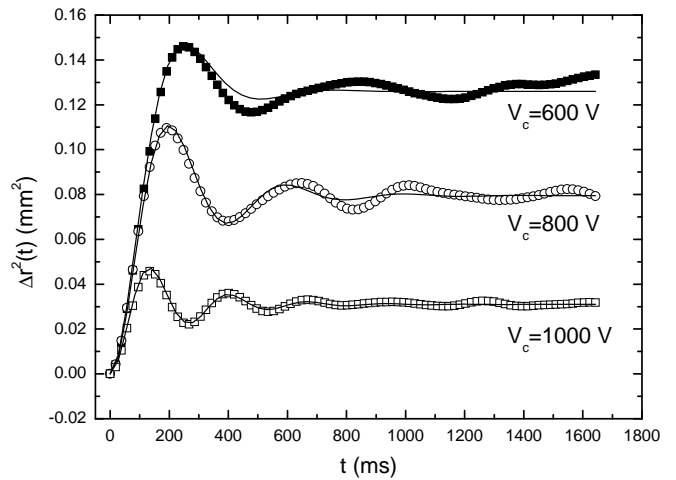}} \caption{Radial
m.s.d. for a single ball in the circular channel for three different
applied voltages. Solid lines show the fits to
equation~\ref{eq:diffusionaussi}.} \label{fig:fig3}
\end{figure}

At low temperatures, the ball oscillates radially with a slow
orthoradial drift. As the temperature increases, the orthoradial
displacements grow, the ball runs over the whole channel with long
free jumps between two sequences of radial oscillations. Note that
no bounces on the channel walls have been observed whatever the
temperature, the ball is only confined by the electrostatic
potential. Its typical shape, determined from the distribution of
radial positions, is presented in Fig.~\ref{fig:fig2}. Whatever
$V_c$, it is parabolic with a stiffness $K$ that increases with the
applied voltage while the mean radial position remains unchanged.
Let us indicate that for too small $V_c$, the electrostatic
confinement looks more like the flat geometrical channel profile
than a parabolic well. Therefore, we will work in the following with
high $V_c$ ($\simeq 1000V$) in order to reinforce the 1D movement.

For a 1D parabolic potential, Langevin equation can be analytically
solved and its characteristic coefficients determined by fits with
the experimental data. The evolution with time of the radial m.s.d.
of a single ball in the channel is shown on Fig.~\ref{fig:fig3}. At
short times, it presents a rapid increase followed by small damped
oscillations and finally reaches a constant value at long times. It
is very well fitted by the theoretical expression obtained through
Langevin formalism in the case of low friction (see the appendix),

\begin{equation}
\Delta r^2(t)=2\langle\langle r^2_0\rangle\rangle\big[1- e^{-\gamma
t/2} \big( \cos(\omega t)+ \frac{\gamma}{2 \omega}\sin(\omega
t)\big)\big], \label{eq:diffusionaussi}
\end{equation}

where $\langle\langle r^2_0\rangle\rangle=k_B T/K$ is the mean
square starting position, $\gamma$ the damping coefficient and
$\omega=\sqrt{\omega_0^2-\gamma^2/4}$ (with $\omega_0=K/m$) the
effective frequency of the oscillations of the ball.

As expected, the larger the applied potential $V_c$ is, the better
the fits are. The frequency $\omega$, which is
temperature-independent and equal to about $20$ Hz within the
explored $V_c$ range, is in accordance with the independently
measured frequency obtained from the Fourier transform of the ball
trajectory, which confirms the validity of the use of the Langevin
equation. The $K$ values determined from these frequencies are in
perfect agreement with those evaluated from the radial distribution
variance and varies as $V_c^2$ as expected from the determination of
the interaction between a ball and the wall~\cite{galatola05}.

From the fits, the damping coefficient $\gamma$ is found to be
independent from $V_c$ and equal to about $10$ s$^{-1}$. This value
is of the same order as the one for free diffusion, which indicates
that the relaxation time $\tau_R=\gamma^{-1}$ results uniquely in
the case of a unique ball from the friction process on the bottom
electrode. Within the temperature range, $\gamma$ is slightly
temperature-dependent but, because of the respective range of
$\omega_0^2$ and $\gamma^2/4$, this dependence is not significant
for $\omega$.

Note that the main feature of the presented fits is that the time
$\tau_R$, which is an important time of reference when analyzing the
orthoradial movement, can be precisely determined for each
experiment.

Another point is the coupling between the orthoradial and the radial
movement in the circular channel. The variations of the angular
m.s.d. for $V_c$ = 1000 V is presented in Fig.~\ref{fig:fig6}. The
long time ($t \gg \tau_R$) behavior is linear, which indicates a
free orthoradial diffusion. This behavior is observed whatever the
temperature. Note the oscillations of the radial m.s.d are not
observed here, which strongly evidences that the orthoradial and
radial movements are independent. This was suggested by the
independence of $K$ with the effective temperature, because
otherwise the effective stiffness would depend on the temperature
through the orthoradial velocity. Moreover this angular free
diffusion time dependence unambiguously confirms that the ball is
only confined by the electrostatic potential and that the balls are
not bouncing on the walls, otherwise different dependencies would be
found~\cite{aslangul99}.

Finally, from the measurement of the diffusion constant of the
orthoradial movement and the measurement of the damping coefficient
through the independent study of the radial movement, one can check
that the fluctuation-dissipation relation is obeyed, which proves
that the thermal equilibrium is reached (as also shown in
Ref.~\cite{coupier05}) and justifies the use of the latter relation
to determine any missing parameters in a given experiment.

Thus the movement of one particle trapped in the circular channel is
twofold. Along the radial direction it is a 1D diffusion in a
parabolic trap whereas it is a free angular diffusion along the
channel. Moreover, this indicates that the formal coupling which
exists between these two coordinates in the Langevin equation, is
not relevant in our experiment. So we can use this set-up in order
to study SFD processes through the orthoradial diffusion of $N$
interacting particles.

\section{Diffusion of interacting particles in a circular channel}
\label{sec:nballs}
\begin{figure}
\resizebox{\columnwidth}{!}{\includegraphics{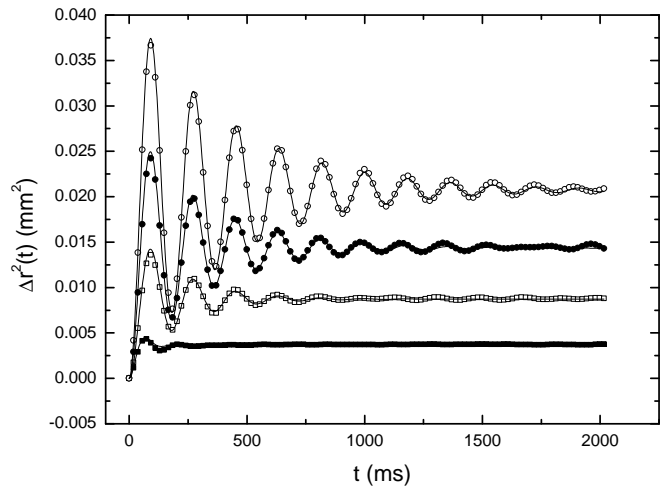}} \caption{Radial
m.s.d of a ball in a 12-ball ring, with $V_c=1000$ V and
$T=3.5\times10^{11}$ K ({\tiny$\blacksquare$}), $7.5\times10^{11}$ K
({\tiny$\square$}), $11.5\times 10^{11}$ K ($\bullet$),
$17.5\times10^{11}$ K ($\circ$).Solid lines shows the fit to
equation~\ref{eq:diffusionaussi}.} \label{fig:fig4}
\end{figure}
\begin{figure}
\resizebox{\columnwidth}{!}{\includegraphics{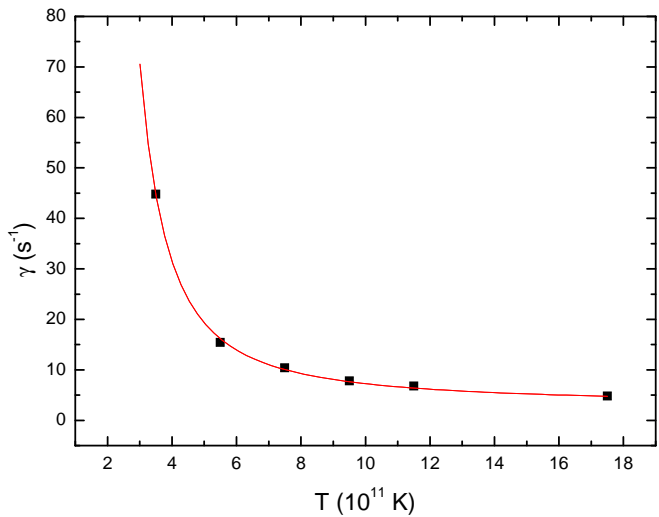}}
\caption{Evolution with the temperature of the damping coefficient
$\gamma$ for $N=12$ balls. Solid line shows the fit to the
Guzman-Andrade law.} \label{fig:fig5}
\end{figure}

We now consider rings of  $N = 12$ and $N=16$ interacting particles.
These two systems were selected because in our channel geometry,
these rings remain circular and do not present "zig-zag"
configurations which could alter the SFD effects~\cite{schweigert96}
. Furthermore, these rings also appear as outer shells in Wigner
islands whose dynamics will be presented in a next
paper~\cite{coupier06}.

The typical radial confining potentials for $N=12$ are shown on
Fig.~\ref{fig:fig2}. As for $N=1$, they are parabolic and
temperature independent and their stiffness increases as $V_c^2$.
However we observe a stiffening with $N$, which is independently
confirmed by the increase of the frequencies obtained from the fit
of the radial m.s.d. by the expression~\ref{eq:diffusionaussi}
(Fig.~\ref{fig:fig4}). The frequencies $\omega\simeq\omega_0$ are
about $30$ s$^{-1}$ for $N=12$ and 16 whereas they are about 20
s$^{-1}$ for $N=1$. This increase of the frequencies with $N$ for a
small number of balls is in accordance with the numerical study
presented in Ref.~\cite{schweigert96}. Consequently, the 1D
character of the angular movement is amplified in the case of $N$
interacting balls. Moreover, this stiffening makes the parabolic
modelization of the radial well all the more exact, and thus the
determination of the damping coefficient $\gamma$ through the radial
m.s.d. fits all the more accurate. As shown on Fig.~\ref{fig:fig5},
$\gamma$, which remains independent of the applied potential $V_c$
and of the number of balls, is a decreasing function of the
temperature. Let us indicate that this last dependence, well fitted
with the law $\gamma=A e^{B/T}$ suggests that our "thermal bath"
looks more like a liquid than a gas for which $\gamma$ increases
with $T$~\cite{commentviscosity}. We underline it does not imply
that the electrostatic coupling between the balls is modulated by
hydrodynamic effects as in Ref.~\cite{lin02}.

\begin{figure}
\resizebox{\columnwidth}{!}{\includegraphics{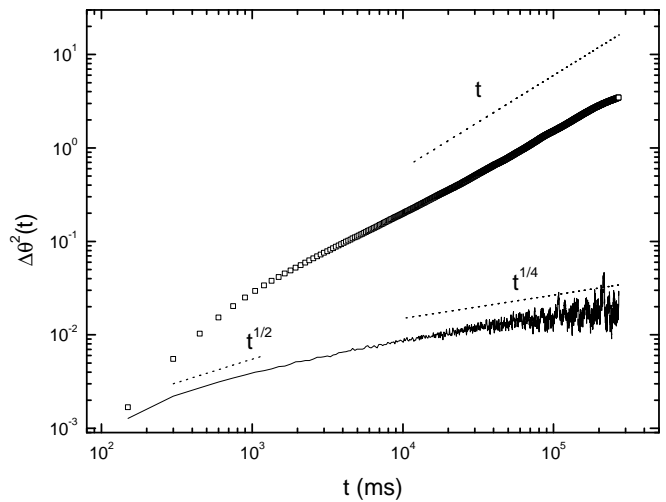}} \caption{Angular
m.s.d. in log scale for a single ball ({\tiny$\square$}, $V_c=1000$
V, $T=15.1\times10^{11}$ K) and a ball embedded in a 12-ball ring
(solid line, $V_c=1000$ V, $T=3.5\times10^{11}$ K). $\theta$ is in
radians.} \label{fig:fig6}
\end{figure}

Along the orthoradial direction, the distribution of the position of
each ball is gaussian with a width proportional to $1/N$. The
amplitude of the displacement is small (about 2 mm), so the
curvature of the channel is not relevant. The variation with time of
the orthoradial m.s.d is shown on Fig.~\ref{fig:fig6} for $N=12$. As
for the single-ball case, no oscillations are observed, confirming
the complete decoupling between the radial and orthoradial movement.
After a $t^2$ increase at short times, a fast transition to a
subdiffusive behavior at long times is observed. However, and it is
the strongest difference with the results presented for colloidal
systems, this increase with time is slower than $t^{1/2}$ at long
times. The same behavior is observed whatever $N$ and $V_c$. This
continuous diffusion slowing down has never been mentioned
previously, all the experiments on colloids concluding in the
existence of only two distinct regimes : free diffusion
characterized by a linear behavior at short times and a $t^{1/2}$
behavior at long times, attributed either to SFD or to hydrodynamics
coupling effects. We can note also that, contrary to the studies on
colloids, no clear linear regime is observed in our experiment.
Indeed, the direct interaction time which marks the switch between
the normal and the subdiffusive regimes and qualitatively
corresponds to the time when a particle "feels" the moving well
around itself is quite short and of the same order as $\tau_R$ which
characterizes the crossover between the $t^2$ and the linear free
diffusion regimes.

Different reasons could explain these differences. The first one
results from the characteristic times of the diffusion itself. In
the colloidal systems, the direct interaction time (about 10 s)
appears on the reported data much larger than $\tau_R$ (which should
be lower than 100 ms), and then the linear regime can appear and be
observed during one decade as in Ref.~\cite{lutz04_2}. By contrast,
in our case, the interaction is probably more long range, thus the
presence of adjacent particles is more rapidly efficient and the
crossover with the subdiffusive regime appears sooner. This
assumption is confirmed by the decrease of the crossing time as the
number of balls increases, in accordance with the enhancement of the
repulsive interaction. Thus the linear regime is drastically
reduced, hidden by the rapid emergence of the subdiffusive behavior.
The second reason may be the limitation in time in the colloids
experiments. In the latter, the subdiffusive behavior is put in
evidence during another decade after the direct interaction time and
is fitted as $t^{1/2}$. However, the m.s.d. time evolution in the
figures presented Ref.~\cite{lutz04_2,wei00} exhibits ultimate
points under the $t^{1/2}$ fit curves which suggests a slight
slowing down beyond. This slowing down could be observed in our
experiments and not in colloidal systems simply because they are
performed up to a time of three order of magnitude higher than our
direct interaction time which is roughly equal to $\tau_R\simeq100$
ms. Longer recording in colloidal systems could give behaviors
similar to ours.

Our results show unambiguously that the behaviors observed for
colloidal systems are not due to hydrodynamic effects but describe
real SFD. However, the fundamental origin of the observed slowing
down is not clear yet. The theoretical results which predict
$t^{1/2}$ at long times focus on infinite systems and have been
performed either for hard-core interactions or for overdamped
systems. Neither our Wigner rings (the oscillations observed in our
radial m.s.d. prove moreover that our damping is weak) nor the
colloidal systems correspond to these assumptions. So the usual
theoretical predictions are not fully adapted to analyze our results
which may not be discussed in this simple frame. We suggest that the
observed slowing down could result from the periodicity of the
system associated with an interaction that would be neither
hard-core nor linear. Indeed, it has been proved that for hard-core
interacting particles, the diffusion is normal for periodic
systems~\cite{vanbeijeren83}. On the other hand, for a linear
interaction, the system of coupled Langevin equations can be solved
in normal coordinates, and the contribution of the mode
corresponding to the collective rotation gives a classical linear
diffusion. Since neither our system nor the colloidal ones exhibit
such a diffusion, this suggests that the distance dependence of the
interaction is of high importance for periodic systems. These
situations could introduce specific correlations between particles,
not described in the usual models. It would be interesting to
develop theoretical models including this new constraint.

\section{Conclusion}
\label{sec:conclusion}

 Single file diffusion of electrostatically
confined and interacting millimetric particles was studied. Their
erratic motion conveyed through the mechanical shaking of the set-up
was proved to be well described by the Langevin equation and thus to
be Brownian. The characteristic parameters of the equation are well
determined by fits of their radial mean square displacements.

Long time behavior of the angular mean square displacement was
explored. We exhibit a subdiffusive behavior with a continuous
slowing down that leads to a slower increase than the $t^{1/2}$
growth observed in colloidal systems. We suggest this might be a
consequence of a non hard-core interaction in a periodic system.

\appendix
\section{The Langevin approach for a trapped particle}
\label{sec:appendix}

Experimental measurement of the time-dependent mean square
displacement of a trapped particle requires to pay attention to the
fact that the initial conditions of the measured trajectories are
not equivalent, therefore the results are not exactly the same as
the one usually presented in the theories. After a short summary of
those usual results, we present what one should expect from an
experimental determination such as ours.

We consider a 1D particle of mass $m$ and coordinate $x(t)$ trapped
in a parabolic well of stiffness $K$ and immersed in a thermal bath
characterized by a friction force $-\alpha \dot{x}$ and a rapidly
fluctuating force $F$ which obeys $\langle F(t)\rangle=0$, where the
notation $\langle\,\rangle$ denotes an ensemble averaging. Newton's
second law of motion gives :
\begin{equation}
\label{eq:langevin} m \frac{d^2x}{dt^2}=-\alpha \frac{dx}{dt}-K x
+F(t).
\end{equation}

The major hypothesis of the Langevin equation is that $F$ is
fluctuating with a characteristic time $\tau_c$ that is much lower
than the time $\tau_R=\gamma^{-1}$ that characterizes the variation
of the speed of the particle, where $\gamma=\alpha/m$ is the damping
coefficient~\cite{tempschar}. Then we can write $\label{eq:moment}
\langle F(t)F(t')\rangle=g\delta(t-t')$. Note that $\gamma$ has the
same microscopic origin as the strength $F$, that's to say the
numerous interactions with the particles of the thermal bath, then
we should expect $\gamma$ to depend also on $g$, therefore on $T$.

If one considers a given initial condition $(x_0,v_0)$ and solves
Eq.~\ref{eq:langevin}, one finds, in the case of a free particle
($K$ =0), the following short and long time behaviors for the
mean-square displacement $\Delta x^2(t)=\langle (x(t)-\langle
x(t)\rangle)^2\rangle$ :
\begin{eqnarray}
\Delta x^2(t)&\underset{t\to0}{\sim}& \frac{g}{3m^2}
t^3\label{eq:freeshort},\\
&\underset{t\to\infty}{\sim}&2 D t,
\end{eqnarray}

where $D=\frac{g}{2m^2 \gamma^2}=\frac{k_B T}{m \gamma}$ is the
diffusion constant.

For a trapped particle, the Langevin equation remains linear and is
easily solved. In the case of weak friction ($\gamma < 2
\omega_0=2\sqrt{K/m}$), one finds, for fixed $(x_0,v_0)$ initial
conditions:
\begin{eqnarray} \lefteqn{\Delta x^2(t)=\frac{g}{2 m^2 \gamma \omega^2 \omega_0^2}\big[\omega^2+e^{-
\gamma t}}\nonumber\\
 &&\times\big(\frac{\gamma^2}{4}
\cos(2 \omega t)-\frac{\gamma \omega}{2} \sin(2 \omega
t)-\omega_0^2\big)\big],
\end{eqnarray}

where $\omega^2=\omega_0^2-\frac{\gamma^2}{4}$.

 In particular, for long times ($t\gg \tau_R$),
\begin{equation}
\Delta x^2 (t)\sim \frac{g}{2 m^2 \gamma \omega_0^2}=\frac{k_B
T}{K}.
\end{equation}

However, in a majority of experiments, the averaging
$\langle\,\rangle$ is obtained by taking different trajectories
considered as different realizations of the random theoretical
trajectory. Those trajectories are obtained by shifting the initial
time, that's to say, if $x(t)$ is the measured trajectory relatively
to a given reference point, one considers the trajectories
$x(t+t_0)-x(t_0)$, where each $t_0$ value gives a new trajectory.

For a free diffusion, this implies, since all points in space are
equivalent, to take trajectories with the initial condition $x_0=0$.
However, initial speed $v_0$ cannot be controlled and summation over
the trajectories will also implies an averaging over the $v_0$
distribution.  If the experiment has started for a time larger than
$\tau_R$, we can consider that this distribution is the stationary
one, thus the first moment of $v_0$ is zero and the second is
$\frac{k_B T}{m}$. Note that this artefact cannot be avoided unless
one is able to start all experiments with exactly the same
conditions. Denoting $\langle\langle\,\rangle\rangle$ this new
averaging, we find $\langle\langle x(t)\rangle\rangle =0$ and,
noting that the $v_0$ and $F$ distributions are independent, the
following behavior for the mean square displacement :
\begin{eqnarray}
\Delta x^2(t)&\underset{t\to0}{\sim}& \frac{k_B T}{m}
t^2\label{eq:freeshortbis},\\
&\underset{t\to\infty}{\sim}&2 D t.
\end{eqnarray}

The main consequence is that $\Delta x^2(t)$ starts as $t^2$ for
short times, and not as $t^3$ (Eq.~\ref{eq:freeshort}). Long time
behavior is not modified.

For a trapped diffusion, not all starting positions are equivalent
and the averaging on the trajectories $x(t)-x_0$ necessarily implies
averaging not only on the random force distribution but also on the
initial condition $(x_0,v_0)$ distribution. Again, if the experiment
has started for a time larger than $\tau_R$, we can consider that
this distribution is the stationary one, then that the first moment
of $x_0$ is zero and the second is  $\frac{k_B T}{K}$.

We find that $\langle\langle x(t)-x_0 \rangle\rangle =0 $ and
\begin{eqnarray}
\lefteqn{\Delta x^2(t)=2\langle\langle x^2_0\rangle\rangle\ }\nonumber\\
&&\times \big[1- e^{-\gamma' t} \big( \cos(\omega t)+
\frac{\gamma'}{\omega}\sin(\omega t)\big)\big]. \label{eq:diffusion}
\end{eqnarray}

In particular :
\begin{equation}
\Delta x^2 \underset{t\to 0}{\sim} \frac{k_B T}{m} t^2.
\end{equation}

 Short time behavior is thus independent from the well
and similar to the free diffusion case. Note also that the infinite
time limit $\frac{2k_BT}{K}$is twice the theoretical value in the
case of identical initial condition trajectories.


\begin{thebibliography}{16}
\bibitem[{tem()}]{tempschar}
\bibinfo{note}{Let us indicate that, in the light of our experiment the
  Langevin formalism seems to apply even if the time $\tau_R$ characterizing
  the relaxation process is not that much larger than the time $\tau_c$ linked
  with the surrounding thermal bath. The latter can be evaluated from the
  average frequency of our "white noise", which indeed is white within the
  range 0-200 Hz. So we have $\tau_c\simeq 10 $ms, to be compared with
  $\tau_R\simeq 100 $ms}.
\end{thebibliography}
\end{document}